\documentclass[11pt]{article}
\usepackage{graphicx,setspace}
 \usepackage[a4paper,
            bindingoffset=0.2in,
            left=1in,
            right=1in,
            top=1in,
            bottom=1in,
          footskip=.5in]{geometry}
\usepackage{amsmath}
\usepackage{xcolor}
\usepackage{hyperref}
\usepackage{natbib}
\usepackage{bm,color}

\title{An Online Algorithm for Bayesian Variable Selection in Logistic Regression Models With Streaming Data}
\author{
Payel Ghosal \footnote{PhD Candidate, University of Wisconsin–Madison}, Shamriddha De \footnote{PhD Candidate, The University of Iowa}, and Joyee Ghosh \footnote{Associate Professor, The University of Iowa \\
{Part of the work presented in this article was completed as Payel Ghosal's MS project at The University of Iowa, under supervision of Joyee Ghosh}}
}

\date{}

\begin{document}
\maketitle

\begin{abstract} 
\noindent   
In several modern applications, data are generated continuously over time, such as data generated from smartwatches. We assume data are collected and analyzed sequentially, in batches. Since traditional or offline methods can be extremely slow, \citet{ghos:tan:luo:2020} proposed an online method for Bayesian model averaging (BMA). Inspired by the literature on renewable estimation, they developed an online Bayesian method for generalized linear models (GLMs) that reduces storage and computational demands dramatically compared to traditional methods for BMA. The method of 
\citet{ghos:tan:luo:2020} works very well when the number of models
is small. It can also work reasonably well in moderately large model spaces. For the latter case, the method relies on a screening stage to identify important models in the first several batches via offline methods. Thereafter, the model space remains fixed in all subsequent batches. In the post-screening stage, online updates are made to the model specific parameters, for models selected in the screening stage. For high-dimensional model spaces, the chance of missing important models in the screening stage is more likely. This necessitates the development of a method, which permits the model space to be updated as new batches of data arrive. In this article, we develop an online Bayesian
model selection method for logistic regression, where the selected model can potentially change throughout the data collection process. We use simulation studies to show that our new method can outperform the method of 
\citet{ghos:tan:luo:2020}. Furthermore, we describe scenarios under which the gain
from our new method is expected to be small. We revisit the traffic crash data analyzed by \citet{ghos:tan:luo:2020} and illustrate that our new model selection method can have better performance for variable selection.
\end{abstract}

\doublespacing

\section{Introduction}  
In the intricate tapestry of modern data dynamics, streaming data occupies an important role with a continuous flow of information at high volumes. This mode of data generation poses a unique set of challenges and opportunities for individuals or organizations storing or analyzing such data.

Traditional or offline methods store the entire data stream of individual records,
and analyze the collective data, after every batch arrives.
An online technique called renewable estimation was proposed by \cite{luo:song:2020} for analyzing streaming data, which does not require storing  
individual records. Instead, it
stores relevant summary statistics for the data in past batches, and updates or renews the estimates from previous batches, after a new batch of data arrives.
So at every step, their online method assumes having access to individual records for the current batch of data, and relevant summary statistics for the historical data.
This leads to computational savings on both storage and 
execution time, compared to offline methods.
The focus of \cite{luo:song:2020} has been on developing renewable estimates of the maximum likelihood estimate (MLE) in GLMs, and in particular they discussed applications 
of their method to linear, logistic, and Poisson regression in the full model. Later \citet{Han:Luo:Lin:Huan:2024} extended the online framework to
high dimensional linear regression and \citet{Luo:Han:Lin:Huan:2023,Ma:Lin:Gai:2023} to high dimensional GLMs via online penalized likelihood methods. 

Among Bayesian methods using the renewable online estimation framework, \citet{Wang:Chen:Schi:Wu:Yan:2016}
provided online Bayesian variable selection methods for linear regression using DIC, in a small model space with $2^4$ models, where the model selection
criterion can be enumerated for all models.
Utilizing the concept of renewable estimation, \citet{ghos:tan:luo:2020} introduced an online BMA method for GLMs. This method enumerates all models for small model spaces. It utilizes offline Markov Chain Monte Carlo (MCMC) sampling for non enumerable model spaces, to select a promising set of models, during a screening stage. In the subsequent post-screening stage, the model space is held fixed, when the online BMA method is applied.

While the proposed method by \citet{ghos:tan:luo:2020} provides a promising solution, challenges persist. Notably, there could be a potential gap in identifying all models with high posterior probability during the screening stage, when offline MCMC sampling is applied. This brings us to the forefront of our research objective: proposing and exploring the utility of  a new online Bayesian model selection method for logistic regression, that effectively addresses the challenges intrinsic to high-dimensional streaming data.

This paper is organized as follows. In Section \ref{sec:bvs} we provide a brief review of BMA in the context of logistic regression models.
In Section \ref{sec:online:fixed}, we set up the notation for 
streaming data and discuss the method of \citet{ghos:tan:luo:2020}. In Section \ref{sec:online:changing} we describe our new method for online Bayesian model selection for logistic regression.
Empirical studies are conducted in Section \ref{sec:sim}
to show the benefits of the new method. A data on
traffic crash accidents is analyzed in Section \ref{sec:real},
in which the new method is shown to have variable selection performance closer to the offline method, than the method of \citet{ghos:tan:luo:2020}. Finally, the main contributions of this article are summarized in Section \ref{sec:summary}.

\section{BMA in Logistic Regression Models} 
\label{sec:bvs}

Consider independent binary response variables $y_i$ and $(p+1)$ dimensional predictor vectors $x_i^T=(x_{i0}=1,x_{i1},\dots,x_{ip})$ for $p$ predictor variables along with the intercept for observations $i=1,2,\dots,n$, generated from the logistic regression model, where $n$ is the sample size.
Combining all the observations on the response and predictor variables, we write the response vector as $y=(y_1,y_2,\dots,y_n)^T$ and the $n\times (p+1)$ design matrix as $X$, whose $i$th row is given by $x_i^T$, $i=1,2,\dots,n$.

Let $\gamma =(\gamma_1,\gamma_2,\dots,\gamma_p)^T$ denote a vector of $p$ indicator variables such that $\gamma_j=1$ indicates the $j^{th}$ predictor variable is included in the model and $\gamma_j=0$ indicates the $j^{th}$ predictor variable is excluded from the model with $p$ covariates. We include the intercept term in each model, and do not perform variable selection on it. The logistic regression model is given as:

$$
y_i|x_i,\beta_\gamma,\gamma \sim {\rm Bernoulli}(\pi_i),
\  i=1,2,\dots,n,
$$
where $l{\rm ogit}(\pi_i)=x_{i\gamma}^T \beta_\gamma$.

Let the prior distributions of the unknown parameters $\beta_\gamma$ and  $\gamma$ be denoted by $p(\beta_\gamma | \gamma)$ and $p(\gamma)$ respectively. 
The posterior distribution over models is given as
\begin{equation}
  p(\gamma|y) = \frac{p(y|\gamma) p(\gamma)}{\sum_\gamma p(y|\gamma) p(\gamma)},  
  \label{eq:postprob}
\end{equation}
where $p(y|\gamma)$ is the marginal distribution of the response vector $y$ under model $\gamma$, with $\beta_\gamma$ integrated out with respect to its prior. When this integral is not available in closed form, $p(y|\gamma)$ is not analytically tractable. However, for a large sample size,  $p(y|\gamma)$ can be approximated by $e^{-\frac{BIC(\gamma)}{2}}$ \citep{Clyde:2002}, where $BIC(\gamma)$ (defined in (\ref{eq:bic})) is the Bayesian information criterion (BIC) for
model $\gamma$.  With the BIC approximation to the marginal likelihood $p(y|\gamma)$, (\ref{eq:postprob}) can be approximated as
\begin{equation}
  p(\gamma|y) = \frac{e^{-\frac{BIC(\gamma)}{2}} p(\gamma)}{\sum_\gamma e^{-\frac{BIC(\gamma)}{2})} p(\gamma)}.  
  \label{eq:postprob.bic}
\end{equation}

The model space contains $2^p$ models. For a small model space, (\ref{eq:postprob.bic}) can be enumerated for all models. However, for large $p$, all the $2^p$ models cannot be stored in computer memory, so typically sampling algorithms are used for sampling promising models from the model space. Therefore, in such cases, a subset of the model space is visited using algorithms such as MCMC. 
Finally, the importance of each predictor variable can be assessed with the marginal posterior inclusion probability. It is given by
$$p(\gamma_j=1| y) = \sum_{\gamma: \gamma_j=1} p(\gamma|y), \ j=1,2,\dots,p.$$ For enumerable model spaces, $p(\gamma_j=1| y)$ can be calculated explicitly. For larger model spaces, it can be estimated using the relative frequency of sampling using MCMC.

\section{Online Method of \citet{ghos:tan:luo:2020}: Fixed Model Selection}\label{sec:online:fixed}
Consider $b$ batches of data $D_1=\{X_1,y_1\},$ $D_2=\{X_2,y_2\},\dots,$ $D_b=\{X_b,y_b\}$, with sample sizes $n_1,n_2,\dots,n_b$ respectively. Let $D_b* = \{D_1,D_2,$ $\dots,D_b\}$ be the datasets aggregated up to batch $b$ and let $N_b=n_1+n_2+\dots+n_b$ denote the corresponding cumulative sample size. For large $N_b$, the calculation of the MLE of 
$\beta$ (under the full model),
denoted by $\hat \beta_b$, 
that maximizes the log-likelihood 
is computationally intensive. In renewable estimation, proposed by \citet{luo:song:2020}, $\hat \beta_b$ is approximated by $\tilde{\beta_b}$, based on summary statistics from historical data $D_{(b-1)}*$ and the individual observations in the current batch of data $D_b$.

Let $l(\beta_\gamma;D_b*)$ denote the log-likelihood up to the $b$th batch of datasets, for model $\gamma$. For the model
$\gamma$, BIC is defined as 
\begin{equation}
BIC(\gamma) = -2l(\hat \beta_\gamma;D_b*) + p_\gamma log(N_b),
\label{eq:bic}
\end{equation}
where $\hat \beta_\gamma$ maximizes the log-likelihood  $l(\beta_\gamma;D_b*)$ when we have data up to batch $b$, and 
$p_\gamma=(\sum_{j=1}^{p}\gamma_{j}+1)$ is the number of regression parameters in model $\gamma$ including the intercept term. While 
$\hat \beta_\gamma$ is expensive to calculate for streaming data,
one can replace it by its renewable estimator, $\tilde \beta_\gamma$, proposed by \cite{luo:song:2020}. However, for implementing BMA, the availability of $\tilde \beta_\gamma$ alone is not enough. One would also need an approximation to the log-likelihood evaluated at the renewable estimator, in order to approximate $BIC(\gamma)$. Without access to individual records in $D_{(b-1)}*$, $l(\tilde \beta_\gamma;D_b*)$ cannot be calculated. Thus \citet{ghos:tan:luo:2020} proposed two 
approximations to $l(\tilde \beta_\gamma;D_b*)$, which they called Online 1 and Online 2. Online 2 is based on a Taylor series expansion that retains more higher order terms than Online 1. In this article we only focus on Online 2, because it is expected to be more accurate than Online 1.

For small model spaces, the Online 2 method can be used to approximate
the marginal likelihood of all models in the model space, which in turn leads to the calculation of online posterior probabilities of models, and BMA estimates of regression coefficients.
For large model spaces, storing all models is computationally infeasible. In this case, \cite{ghos:tan:luo:2020} used offline MCMC during an initial screening stage, and monitored a metric to determine the stability of estimates across successive batches. When this metric reached an acceptable threshold, or the aggregate sample size was already too large, screening was stopped, and the subset of models visited by MCMC was used for online computation of model specific regression coefficients for all remaining batches. 

While there is some flexibility regarding the choice of the metric, in the
empirical studies the authors used a metric based on the estimates of the class
probabilities, $p(y_{ki}=1 | D_b*), \ k=1,2.\dots,b, \ i=1,2, \dots, n_k$, from two successive batches. The metric used the following RMSE between the estimates from batches $b$ and 
$b-1$:
\begin{equation}
\sqrt{\frac{1}{\sum_{k=1}^{b} n_k}\sum_{k=1}^{b}\sum_{i=1}^{n_k}{\left(\frac{\exp(x_{ki}^T\hat{\beta}_{BMA;b})}{1+\exp(x_{ki}^T\hat{\beta}_{BMA;b})} -\frac{\exp(x_{ki}^T\hat{\beta}_{BMA;b-1})}{1+\exp(x_{ki}^T\hat{\beta}_{BMA;b-1})}\right)}^2},
\label{eq:metric}
\end{equation}
where $\hat{\beta}_{BMA;b-1}$ and $\hat{\beta}_{BMA;b}$ are BMA estimates of $\beta$ from offline MCMC runs after data arrive from batches $b-1$
and $b$ respectively. 
\citet{ghos:tan:luo:2020} used 
a threshold of 0.02 to check stability. When the metric in (\ref{eq:metric}) reached a value equal to or less than the threshold, or the number of batches was 10, the screening period ended, based on whichever happened first.


 In particular, the offline MCMC in the screening stage was implemented based on the MC$^3$ algorithm \citep{Raftery:et.al.:1997}. 
This is a Metropolis--Hastings algorithm, where a new model $\gamma^*$ is proposed by randomly choosing one of the components of the current model
$\gamma$, and changing the component from 0 to 1 or from 1 to 0. The proposed model is accepted with a probability that equals ${\rm min}\{1,\frac{p(\gamma^*|y)}{p(\gamma|y)}\}$.

\section{A New Online Algorithm: Changing Model Selection} \label{sec:online:changing}


 The primary essence of the approach explicated in \cite{ghos:tan:luo:2020} in the previous section lies exclusively in its utilization of an \textbf{offline} MCMC algorithm during an initial screening stage, to sample models from promising regions of the model space. Nonetheless, realistic scenarios may arise when the MCMC in the screening stage fails to identify some important models. Since the method of \citet{ghos:tan:luo:2020}
 does not sample any new models in the 
 post-screening stage, but only updates the previously sampled models based on new batches of data, it is unable to recover this lost information. In such cases, an online algorithm that can potentially sample new models in every batch, without storing individual records
 from past batches, is highly desirable.



\subsection{An Approximate Linear Model for Online Estimation of Posterior Model Probabilities}

In this article, we introduce a 
solution to tackle the aforementioned challenges. For the 
Bayesian logistic regression model, implementing an online algorithm which allows selected models to change across batches is difficult, without access to individual records. The novelty in our proposed method is to utilize
a fully online approximate approach employing linear models. Our method approximates the posterior probabilities of logistic regression models in BMA, via a linear model approximation, in a computationally feasible manner. 
In the following, we first describe the online updating procedure for the full linear regression model with all predictors in Section \ref{sec:lmfull}, and
suppress the notation $\gamma$ (which denotes models)
for simplicity. Thereafter, we explain 
how to extend the method to
different models via online MCMC sampling in Section \ref{sec:lmonline}.

\subsubsection{Online Estimation Under the Full Linear Model}
\label{sec:lmfull}

Consider the linear model: 
\begin{equation}
   {y}_{b}  = X_b \alpha + {\epsilon_b}, 
   \label{eq:lm1}
\end{equation}
where $y_b$ and $X_b$ are the $n_b \times 1$ vector of response variables, and the $n_b \times (p+1)$ design matrix, respectively, from the $b^{th}$ batch of data with sample size $n_b$, $\alpha$ is the $(p+1) \times 1$ vector of linear regression coefficients with the intercept term, and $\epsilon_b$ is the $n_b \times 1$ vector of random errors. Under the above linear model, we assume 
\begin{equation}
y_{bi}|x_{bi},\alpha \sim N( x_{bi}^T \alpha,\sigma^2), \text{ independently  } \forall i\in \{1,2,\dots,n_b\},
\label{eq:lm2}
\end{equation}
with an unknown error variance $\sigma^2$. As the response variables are binary, this is not an ideal mathematical model. This model is chosen for its tremendous computational 
advantage over the logistic regression model, in a high-dimensional setting with streaming data.
In Section \ref{sec:sim} we will illustrate via empirical studies that this model can be quite informative about the
importance of variables.

As mentioned in Section \ref{sec:online:fixed}, the streaming data up to the $b^{th}$ batch is denoted by $D_b^*=\{D_1=\{X_1,y_1\},D_2=\{X_2,y_2\},\dots,D_b=\{X_b,y_b\}\}$. Therefore the sample size up to the $b^{th}$ batch is given by $N_b=(n_1+n_2+\dots+n_b)$, and the corresponding combined design matrix $X_b^*$, and the vector of combined response variables $y_b^*$, up to $b$ batches are given by,
\[
    D_b^* = \begin{Bmatrix}
    X_b^* =
    \begin{bmatrix}
           X_{1} \\
           X_{2} \\
           \vdots \\
           X_{b}
         \end{bmatrix}, \
y_b^* = \begin{bmatrix}
           y_{1} \\
           y_{2} \\
           \vdots \\
           y_{b}
         \end{bmatrix} 
     \end{Bmatrix}.    
\]
The corresponding log-likelihood under (\ref{eq:lm2}), up to the $b^{th}$ batch is given by
\begin{equation}
    l(\alpha,\sigma^2;D_b^*) = -\frac{N_b}{2} \log(2\pi) - \frac{N_b}{2} \log(\sigma^2) - \frac{1}{2\sigma^2} (y_b^{*T}y_b^* - 2\alpha^T X_b^{*T} y_b^*+ \alpha^T X_b^{*T}X_b^* \alpha), 
\label{eq:loglik.batch.b}
\end{equation}
where $\alpha$ and $\sigma^2$ are unknown parameters. The log-likelihood up to the $b^{th}$ batch of data, evaluated at the MLE of the unknown parameters can be expressed as:
\begin{equation}
 l(\hat \alpha_{b}^{*}, \hat \sigma_{b}^{*2};D_b^*) = -\frac{N_b}{2} \log (2\pi) - \frac{N_b}{2} \log(\hat \sigma_{b}^{*2}) - \frac{1}{2\hat \sigma_{b}^{*2}} (y_b^{*T}y_b^* - \hat \alpha_{b}^{*T} X_b^{*T} y_b^*), 
\label{eq:loglik.batch.b.mle}
\end{equation}
where the MLE of $\alpha$ and $\sigma^2$, up to the data for $b$ batches, are given as
\begin{equation}
  \hat \alpha_{b}^{*} = (X_b^{*T} X_b^*)^{-1}X_b^{*T} y_b^*, \ 
\hat \sigma_{b}^{*2} = \frac{y_b^{*T}y_b^*-\hat \alpha_{b}^{*T} X_b^{*T} y_b^*}{N_b}. 
\label{eq:mle.batch.b}
\end{equation}
If individual records are stored, that is $X_b^*$ and $y_b^*$ are stored for each batch of data, the above computation can be done easily, which we refer to as the offline or traditional method for computation. 

For online computing, we propose an alternative method for recovering the log-likelihood at the MLE. We initiate the process with the first batch, $b=1$, and compute the corresponding log-likelihood at the MLE as follows,
$$
l(\hat \alpha_{1}^{*}, \hat \sigma_{1}^{*2};D_1^*) = -\frac{N_1}{2} \log(2\pi) - \frac{N_1}{2} \log(\hat \sigma_{1}^{*2}) - \frac{N_1}{2},
$$
where $D_1^*=D_1$, $y_1^* = y_1, X_1^* =X_1,N_1=n_1$, and
the MLE of $\alpha$ is $\hat \alpha_{1}^{*}
= (X_1^{T} X_1)^{-1}X_1^{T} y_1$, and that of $\sigma^2$ is
$\hat \sigma_{1}^{*2}
= \frac{y_1^{T}y_1-\hat \alpha_{1}^{T} X_1^{T} y_1}{n_1}$. Rather than storing the 
$n_1 \times (p+1)$ matrix of individual records from the first batch, we only store three summary statistics:
$y_1^{*T}y_1^*, X_{1}^{*T}y_{1}^*$ and $X_{1}^{*T}X_{1}^*$, whose dimensions are $1\times 1, (p+1)\times 1$ and $(p+1) \times (p+1)$, respectively. The dimensions of these summary statistics do not grow as the number of batches increases, unlike the $N_b \times (p+1)$ dimensional cumulative data matrix of individual records, up to batch $b$.
Upon arrival of the second batch $D_2=\{X_2,y_2\}$, our approach advocates updating and storing the summary statistics, and then discarding the individual records from the second batch. These updates are expressed as follows:

\begin{align*}
    y_2^{*T}y_2^*= y_1^{*T}y_1^* + y_2^{T}y_2,\\
    X_{2}^{*T}y_{2}^* = X_{1}^{*T}y_{1}^* + X_2^Ty_2,\\
    X_{2}^{*T}X_{2}^* = X_{1}^{*T}X_{1}^* + X_2^TX_2.
\end{align*}
By plugging the above updated summaries in (\ref{eq:mle.batch.b}) and 
(\ref{eq:loglik.batch.b.mle})
with combined sample size $N_2=n_1+n_2$, we get,

$$
l(\hat \alpha_{2}^{*}, \hat \sigma_{2}^{*2};D_2^*=\{D_1,D_2\}) = -\frac{N_2}{2} \log(2\pi) - \frac{N_2}{2} \log(\hat \sigma_{2}^{*2}) - \frac{N_2}{2}.
$$ 

In general, if the matrices are available up to batch $b \in \{1,2,\dots \}$, then these can be updated by the raw data of the $(b+1)^{th}$ batch, i.e., $D_{(b+1)}=\{X_{(b+1)},y_{(b+1)}\}$ without storing the whole data from the past $b$ batches using the following scheme:
\begin{align*}
    y^{*T}_{(b+1)}y_{(b+1)}^* = y^{*T}_{b}y_{b}^* + y_{(b+1)}^Ty_{(b+1)},\\
    X_{(b+1)}^{*T}y_{(b+1)}^* = X_{b}^{*T}y_{b}^* + X_{(b+1)}^Ty_{(b+1)},\\
    X_{(b+1)}^{*T}X_{(b+1)}^* =  X_{b}^{*T}X_{b}^* + X_{(b+1)}^TX_{(b+1)}.
\end{align*}
Evidently for any batch, preserving the above three summaries from the previous batches is enough to compute the log-likelihood and BIC of the linear regression model. 

\subsubsection{Online MCMC Algorithm for Model Selection Under the Linear Model}
\label{sec:lmonline}
\citet{ghos:tan:luo:2020} used the offline MC$^3$ algorithm for logistic regression in the screening stage. In the MC$^3$ algorithm, a new model $\gamma=(\gamma_1,\gamma_2,\dots,\gamma_p)^T$ is proposed at random in every iteration, and is accepted according to the appropriate Metropolis--Hastings ratio defined in Section \ref{sec:online:changing}. 
The challenge of implementing an online version of this MCMC algorithm for high-dimensional logistic regression arises from the fact that the historical summary for a model that has never been sampled in previous batches is not available. Thus, neither the Metropolis--Hastings acceptance probability can be computed, nor do we have estimates of regression coefficients under this new model.

In the context of our approximate linear regression model, we will also use the MC$^3$ algorithm.
The beauty of using the linear model is that a fully online MC$^3$ algorithm can be implemented, which gives results that are mathematically equivalent to the offline MC$^3$ algorithm for linear regression. This is possible because
everything needed for the  MC$^3$ algorithm in a linear model can be expressed via summary statistics, which we describe below.

 At the $b$th batch, the vector $X_{b(\gamma)}^{*T}y_{b}^*$ and the matrix $X_{b(\gamma)}^{*T}X_{b(\gamma)}^*$ can be determined for the model $\gamma$, simply by extracting the entries corresponding to the non-zero components in $\gamma$, from $X_{b}^{*T}y_{b}^*$ and $X_{b}^{*T}X_{b}^*$, respectively. Clearly the log-likelihood evaluated at the MLE under model $\gamma$ can then be easily calculated, which in turn can be used to calculate the BIC for this model. Each proposed model is accepted based on the Metropolis--Hastings acceptance ratio for the MC$^3$ algorithm, defined in Section \ref{sec:online:fixed}. 
 Note that this is now calculated based on the BIC approximation to the marginal likelihood under a linear model, unlike a logistic regression model in \citet{ghos:tan:luo:2020}. 
 
 Henceforth, up to the $b^{th}$ batch, the $\hat \alpha_\gamma$s are available for each of the models $\gamma$, accepted by the MC$^3$ algorithm. A BMA estimate can be computed based on $\hat \alpha_\gamma$s for the accepted models, weighted by the MCMC estimates of posterior probabilities of the models. 
However, as the logistic regression model is approximated by the linear model, the estimate 
$\hat \alpha_{BMA} = (\hat \alpha_{BMA(0)} ,\hat \alpha_{BMA(1)}$
$,\dots,\hat \alpha_{BMA(p)} )^T$ is not compatible with the BMA estimate of the original parameter $\beta$ under logistic regression. 
We experimented with an approach where we transformed the estimated linear model parameters, so that the interpretation of the transformed parameters is compatible with that under the logistic regression model.
However, this estimator exhibited a small discrepancy with the true parameter, which did not always decrease with increasing the sample size. So we adopt a different method to estimate the regression parameters under the logistic regression model, as outlined below.

\subsection{Online Estimates of Logistic Regression
Parameters Under the Median Probability Model}

Ideally, we would like to recover the offline BMA estimate of the parameters of the logistic regression coefficients,
that is approximate the estimate 
$\hat \beta_{BMA} = (\hat \beta_{BMA(0)} ,\hat \beta_{BMA(1)}$
$,\dots,\hat \beta_{BMA(p)} )^T$, via an online method. It turns out that this goal
is quite challenging for logistic regression, in an online setting. So instead of approximating the BMA estimate of $\beta$, we propose a two-step approach for online estimation of $\beta$, under the median probability model (MPM) of \cite{Barb:Berg:2004}. 

The MPM is defined
as the model which includes all predictors
with marginal posterior inclusion probability greater than or equal to half. Let $\gamma_{MPM}$ denote the MPM, then we have $\{\gamma_{MPM(j)} = 1\} \equiv \{p(\gamma_j=1|y) \geq 0.5\}$ and 
$\{\gamma_{MPM(j)} = 0\} \equiv
\{p(\gamma_j=1|y) < 0.5 \}, \ j=1,\dots p$. Based on simulation studies in Section \ref{sec:sim}, we find that $p(\gamma_j=1|y)$ under the logistic regression model are quite accurately estimated by the corresponding estimates under the linear model approximation outlined above, when the sample size is large. 

In the first step, we use $p(\gamma_j=1|y)$, which are estimated via MCMC under the linear model, to determine the MPM under the logistic regression model. Note that the offline estimates of the logistic regression parameters under the MPM cannot be computed, without having access to individual 
records for past batches. To circumvent this problem, we keep track of the online logistic regression estimates under the full model with all predictors, say $\hat{\beta}_{Full}$, for all batches. While keeping track of such online estimates for all models is impossible in a high-dimensional setting, it is straightforward to implement for just one model, namely the full logistic regression model.
Now, in step 2 of the two-step approach, we propose to estimate 
$\hat{\beta}_{MPM}$ under the logistic regression model by setting
$\hat{\beta}_{MPM(j)}=\hat{\beta}_{Full(j)}$ if $\gamma_{MPM(j)}=1$, and  $\hat{\beta}_{MPM(j)}=0$ if $\gamma_{MPM(j)}=0$, for $j=1,2,\dots,p$, and the intercept is always assumed to be in the model. In a nutshell, we use the online linear model approximation to estimate the MPM under the logistic regression model, and then use the selected predictors to form an estimate of $\beta$ based on the online logistic regression estimate from the full model.

\section{Simulation Study}
\label{sec:sim}

 We conducted a simulation study to demonstrate the enhancements brought about by the online algorithm with changing model selection, proposed in Section \ref{sec:online:changing},
 in comparison with existing approaches. Our dataset involves a total of $p=80$ covariates, resulting in $p+1=81$ unknown parameters, including the intercept term. Three simulation scenarios are considered. In the first scenario, the true regression parameters are set as $\beta=(0.1,\dots,0.1,0,\dots,0)^T$, with a value of 0.1 for the initial 21 coefficients and 0 for the successive ones. Consequently, the vector of inclusion indicators $\gamma=(\gamma_1,\dots,\gamma_p)^T$ (which does not include an indicator for the intercept) for the true data generating model exhibits a pattern of $(1,\dots,1,0,\dots,0)^T$ with 20 consecutive occurrences of 1, followed by 60 occurrences of 0.
 
 In the second and third scenarios, the aforementioned vector 
 $\beta$  is multiplied by 1.5 and (1/1.5) to encompass scenarios with stronger and weaker signals, compared to the base scenario. The simulated dataset comprises of $B=50$ batches, with a batch size of $n_b=300$ for each batch. The predictor vector for the $i^{th}$ observation is given by $x_i^T = (x_{i0},x_{i1},\dots,x_{i80})$, where the intercept is included as $x_{i0}=1$. In each batch, the predictors $(x_{i1},x_{i2},\dots,x_{i80})^T$ are generated from independent standard normal distributions for each $i \in \{1,2,\dots,n_b=300\}$. The corresponding responses are generated from independent Bernoulli trials with probabilities determined by a logistic regression model. This whole simulation is replicated $J=25$ times. For prediction, we created a test dataset by generating a new set of predictor variables and response variables from the same model with sample size $N=15,000$.

 The present study compares the following three methods, the first two of which are close competitors, while the third one is the proposed method.
 \begin{itemize}
     \item \textbf{Offline:}
     This approach represents the scenario where the entire dataset from each batch is retained, and offline MCMC is implemented on the aggregated dataset, after each new batch of data arrives. Regrettably, the execution of this approach is nearly infeasible for streaming data due to severe demands on time and storage. Nonetheless, this method serves as the benchmark for comparing alternative methodologies and presents an idealized case for assessment.

     \item \textbf{Online Fixed Model Selection \citep{ghos:tan:luo:2020}:}
     As delineated in Section \ref{sec:online:fixed}, this method performs model selection at an initial screening stage using offline MCMC, and executes online updates for this fixed model space for all subsequent batches. The screening stage is typically confined to a maximum of 10 initial batches of data and uses a threshold of 0.02 for the metric in \eqref{eq:metric}. For brevity and clarity, we refer to this configuration as ``Online Fixed ModelSel''.

     \item \textbf{Online Changing Model Selection:}
     Hereinafter referred to as ``Online Changing ModelSel'', this is our proposed method which only retains $y^Ty$, $X^Ty$, and $X^TX$ from the preceding batches, thereby obviating the need to preserve the entire dataset. The present technique distinguishes itself from Online Fixed ModelSel by updating the model selection after every batch using the MPM, rather than relying upon a fixed model space obtained through a screening phase. The details have been elucidated in Section \ref{sec:online:changing}.
 \end{itemize}

 For the analysis, whenever running an MCMC algorithm is necessary, it is run for $T=10,000$ iterations, after discarding a burn-in of $2,000$ samples. The screening period for Online Fixed ModelSel is set to be 10 batches as in \cite{ghos:tan:luo:2020}, when it is assumed that access to individual records is available. We also take advantage of this fact while obtaining $\hat{\beta}_{MPM}$ in the proposed Online Changing ModelSel method, where we estimate the offline MLE up to batch 10 under the full model. For subsequent batches, we use the R package {\tt RenewGLM} \citep{luo:song:2020} to perform renewable estimation for online updates of the MLE under the full model.

 To facilitate a comprehensive and conspicuous sense of comparison among the three considered methods, we consider the following aspects.
 \begin{itemize}
     \item \textbf{Estimation of regression coefficients:}
     As stated earlier, the regression coefficients are estimated using $\hat{\beta}_{MPM}$ in Online Changing ModelSel. On the other hand, the BMA estimates of $\beta$, denoted as $\hat{\beta}_{BMA}$, are used for both Offline and Online Fixed ModelSel. The comparison involves assessing the root mean squared errors of the estimates with respect to the true $\beta$ in the simulation.

     \item \textbf{Variable selection:}
     To study the efficacies of the methods in variable selection, we look at the vector of inclusion indicators. To estimate any inclusion indicator, the MPM is used, after estimating the corresponding marginal inclusion probability by the proportion of times the said indicator assumes the value 1 in the MCMC sampling. Subsequently, the true positive rates (TPRs) and false positive rates (FPRs) are determined and compared for the three methods. In particular, the TPR is calculated as the proportion of covariates correctly included by the MPM, among all the true signal covariates, while the FPR is computed as the proportion of covariates incorrectly selected by the MPM, among all the true noise covariates.
     
     \item \textbf{Out of sample prediction:}
     The predictive performances of the methods are assessed by evaluating the area under the receiver operating characteristics (ROC) curve, commonly abbreviated as AUC. Typically, a larger value of AUC is preferable for a classifier \citep{Fawc:2006}. For our purpose, we use the R package {\tt pROC} to calculate the AUC values.
     
     \item \textbf{Running time:}
     Based on a single replicate of 50 batches, we compare the running times of the methods, since it is one of the two most important facets intended to be reduced by the online methods. The other important characteristic of online methods is the reduced need for data storage.
 \end{itemize}

\subsection{Simulation Scenarios} 

\subsubsection{Simulation Scenario 1}
\begin{figure}[ht]
    \includegraphics[height=4.4in,width=4.8in]{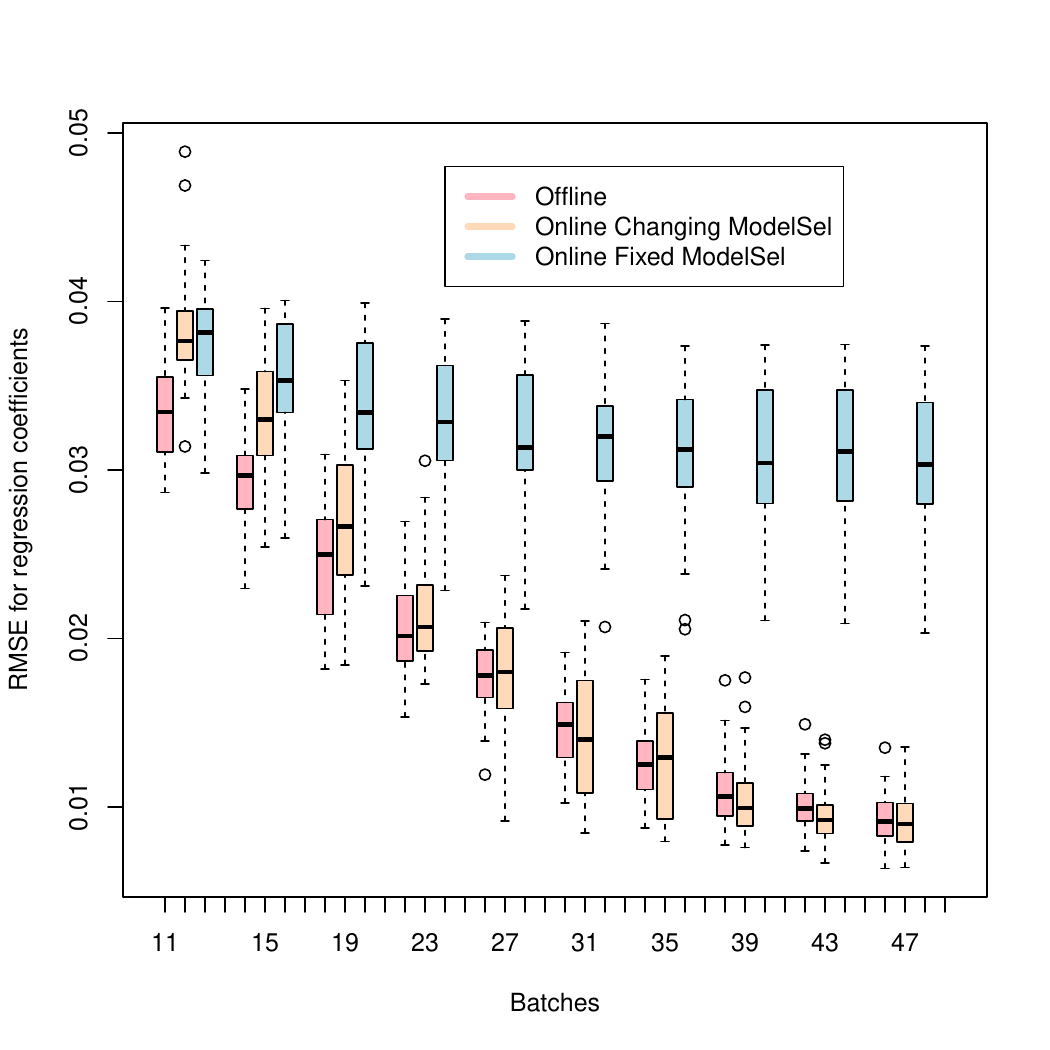}   \includegraphics[height=4.4in,width=4.8in]{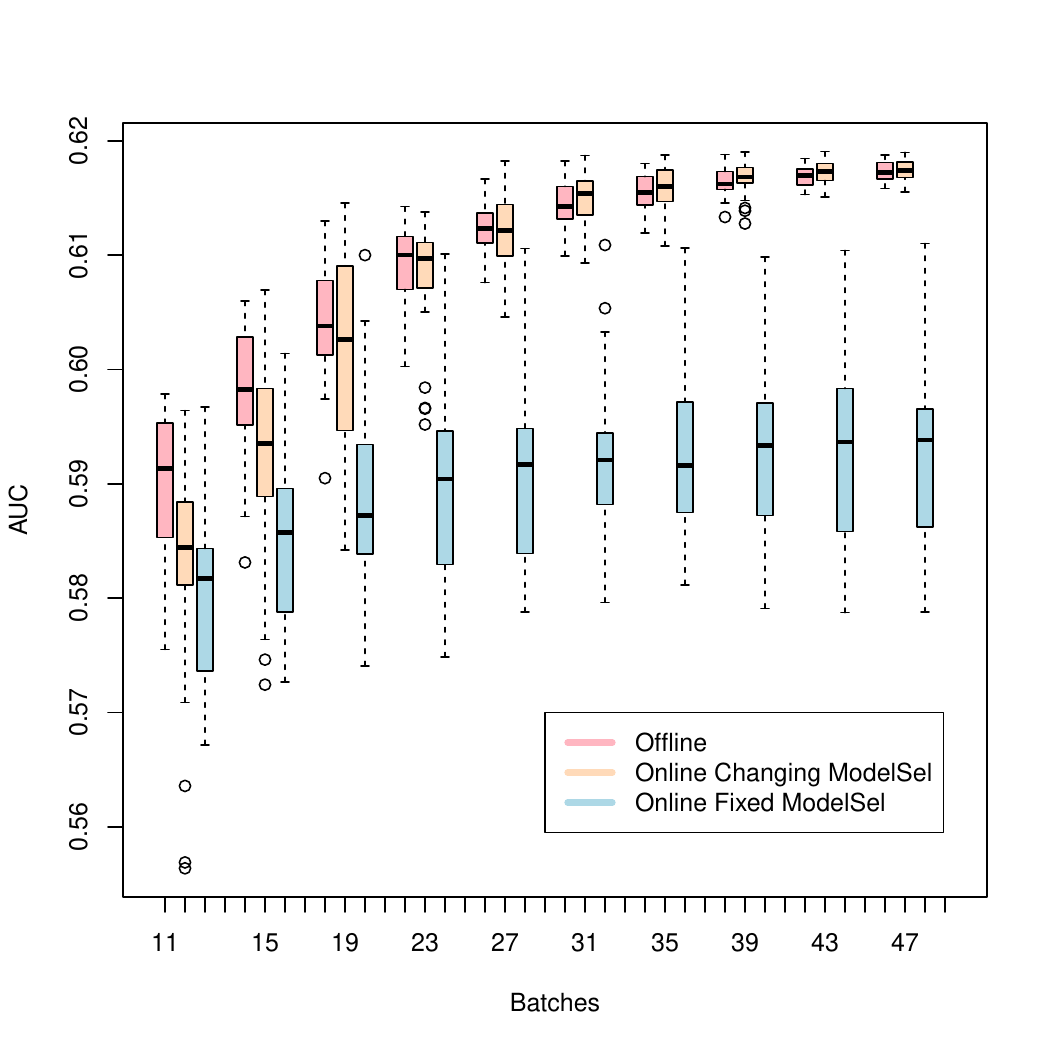}
    \caption{RMSE of regression coefficients and AUC for test data
    for Offline, Online Changing ModelSel, and Online Fixed ModelSel for Scenario 1. The boxplots are based on 25 replicates. \textbf{The running times of Offline and Online Fixed ModelSel are approximately 441 and 16 times that of Online Changing ModelSel for a single replication.}}
    \label{fig:1}
\end{figure}

For the first choice of 
$\beta=(0.1,\dots,0.1,0,\dots,0)^T$,
the plots in Figure \ref{fig:1} reveal distinct performances across the three methods for each of the two metrics. Offline consistently demonstrates low RMSE for regression coefficients, indicating robust estimation accuracy and minimal variability. Online Changing ModelSel similarly shows relatively low RMSE as the sample size increases, underscoring its efficiency in real-time model selection by closely following Offline results. In contrast, Online Fixed ModelSel exhibits higher RMSE, suggesting less effectiveness compared to the other two methods. The values of AUC
for Offline and Online Changing ModelSel become nearly identical for later batches, while Online Fixed ModelSel lags behind. 
Overall, Online Changing ModelSel and Offline outperform Online Fixed ModelSel for this choice of $\beta$, with our new online method offering significant time savings without sacrificing accuracy.

Tables \ref{tab:sim1:tpr} and \ref{tab:sim1:fpr} of TPR and FPR for the first choice of $\beta$ highlight the effectiveness of each method in identifying the true variables
and discarding noise variables. Offline and Online Changing ModelSel consistently identify the true variables with high accuracy. Online Fixed ModelSel exhibits lower TPR and higher FPR. 

All the above results indicate that the Online Fixed ModelSel can fail to identify important predictors or drop irrelevant ones, for the first choice of $\beta$. However, examining the other two choices of $\beta$ is also crucial for assessing the robustness and generalizability of the methods across different signal strengths. It helps to identify potential weaknesses that might not be apparent with just one choice of $\beta$ and ensures that the conclusions are not specific to a single scenario. By analyzing multiple choices of $\beta$, we can validate the findings, providing a more comprehensive comparison of the methods and making informed recommendations for their use in varying data contexts.

\begin{table}[ht]
\centering
\footnotesize{
\begin{tabular}{|c|cccccccccc|}
  \hline
  & 11 & 15 & 19 & 23 & 27 & 31 & 35 & 39 & 43 & 47 \\ 
  \hline
Offline & 0.498 & 0.638 & 0.760 & 0.852 & 0.916 & 0.964 & 0.970 & 0.992 & 0.996 & 0.998 \\ 
   Online Changing ModelSel & 0.508 & 0.636 & 0.756 & 0.852 & 0.914 & 0.964 & 0.972 & 0.990 & 0.998 & 0.998 \\ 
Online Fixed ModelSel  & 0.378 & 0.494 & 0.546 & 0.572 & 0.586 & 0.598 & 0.606 & 0.614 & 0.618 & 0.618 \\ 
   \hline
\end{tabular}
}
\caption{True Positive Rates for Offline, Online Changing  ModelSel, and Online Fixed ModelSel for Simulation Scenario 1 for batches $11,15, \dots, 47$. The values are averaged over 25 replications.}
 \label{tab:sim1:tpr}
\end{table}

\begin{table}[ht]
  \centering
  \footnotesize{
  \begin{tabular}{|c|cccccccccc|}
  \hline
    & 11 & 15 & 19 & 23 & 27 & 31 & 35 & 39 & 43 & 47 \\ 
 \hline
   Offline & 0.005 & 0.007 & 0.005 & 0.003 & 0.003 & 0.005 & 0.003 & 0.002 & 0.003 & 0.003 \\ 
   Online Changing ModelSel & 0.005 & 0.007 & 0.005 & 0.003 & 0.003 & 0.004 & 0.003 & 0.002 & 0.002 & 0.003 \\ 
  Online Fixed ModelSel& 0.003 & 0.003 & 0.005 & 0.007 & 0.011 & 0.013 & 0.019 & 0.022 & 0.025 & 0.025 \\ 
  \hline
  \end{tabular}
  }
\caption{False Positive Rates for Offline, Online Changing ModelSel, and Online  Fixed ModelSel for Simulation Scenario 1 for batches $11,15, \dots, 47$. The values are averaged over 25 replications.}
 \label{tab:sim1:fpr}
  \end{table}
  
\subsubsection{Simulation Scenario 2}
\begin{figure}[ht]
    \includegraphics[height=4.4in,width=4.8in]{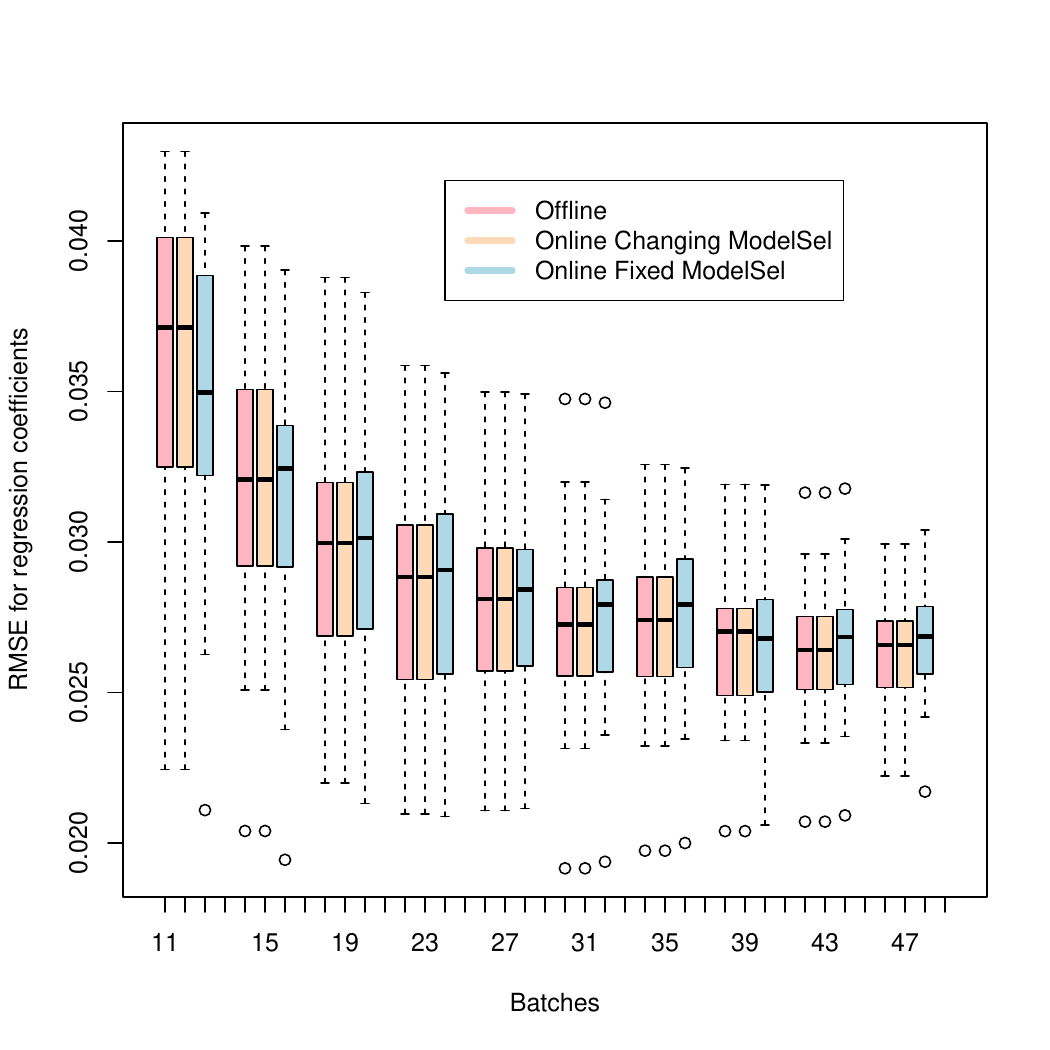}   
    \includegraphics[height=4.4in,width=4.8in]{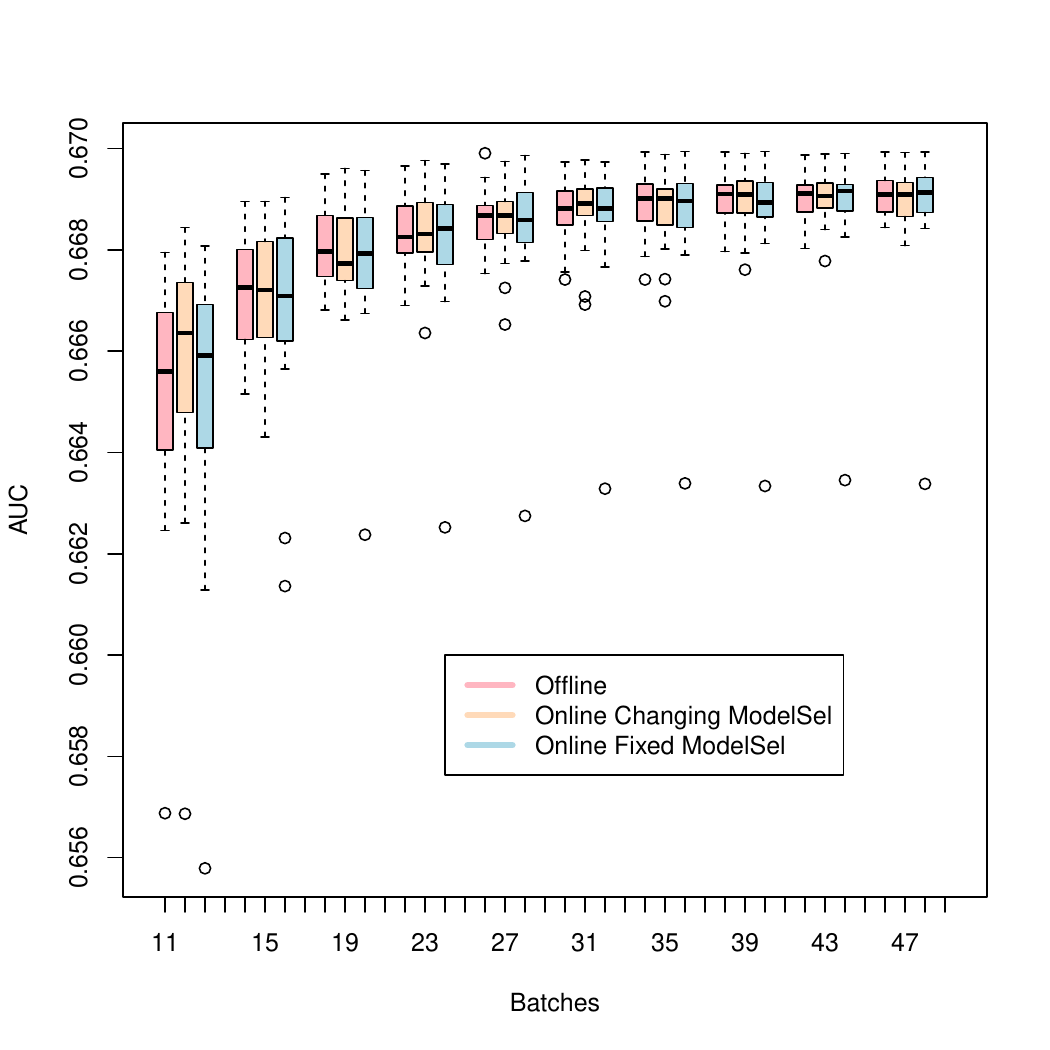}
    \caption{RMSE of regression coefficients and AUC for test data
    for Offline, Online Changing ModelSel, and Online Fixed ModelSel for Scenario 2. The boxplots are based on 25 replicates. \textbf{The running times of Offline and Online Fixed ModelSel are approximately 489 and 12 times that of Online Changing ModelSel for a single replication.}}
    \label{fig:2}
\end{figure}

For the second choice of $\beta=1.5\times(0.1,\dots,0.1,0,\dots,0)^T$,
the plots in Figure \ref{fig:2} and Tables \ref{tab:sim2:tpr} and \ref{tab:sim2:fpr} reveal very similar performances across the three methods and three metrics. All methods have lower RMSE, higher AUC, high TPR, and low FPR, compared to Scenario 1, due to stronger signals. 
Online Fixed ModelSel, while still performing adequately, shows slightly  larger RMSE and more variability in AUC compared to the other two methods, even at batch 47. The 
TPR  and FPR across all methods are quite high and low respectively, reflecting their ability to correctly identify the relevant variables and drop noise variables. 
Overall, while all methods perform well in this scenario due to stronger signals, Offline and Online Changing ModelSel show slightly better and more reliable results, particularly in terms of stability. 

\begin{table}[htb]
\centering
\footnotesize{
\begin{tabular}{|c|cccccccccc|}
  \hline
  & 11 & 15 & 19 & 23 & 27 & 31 & 35 & 39 & 43 & 47 \\
  \hline
Offline & 0.995 & 1.000 & 1.000 & 1.000 & 1.000 & 1.000 & 1.000 & 1.000 & 1.000 & 1.000 \\ 
  Online Changing ModelSel  & 0.990 & 1.000 & 1.000 & 1.000 & 1.000 & 1.000 & 1.000 & 1.000 & 1.000 & 1.000 \\ 
 Online Fixed ModelSel & 0.936 & 0.995 & 0.998 & 0.998 & 0.998 & 0.998 & 0.998 & 0.998 & 0.998 & 0.998 \\ 
   \hline
\end{tabular}
}
\caption{True Positive Rates for Offline, Online Changing  ModelSel, and Online Fixed ModelSel for Simulation Scenario 2 for batches $11,15, \dots, 47$. The values are averaged over 25 replications.}
 \label{tab:sim2:tpr}
\end{table}

\begin{table}[h!]
\centering
\footnotesize{
\begin{tabular}{|c|cccccccccc|}
  \hline
    & 11 & 15 & 19 & 23 & 27 & 31 & 35 & 39 & 43 & 47 \\
  \hline
Offline & 0.006 & 0.006 & 0.003 & 0.001 & 0.001 & 0.002 & 0.003 & 0.004 & 0.004 & 0.005 \\ 
  Online Changing ModelSel & 0.006 & 0.006 & 0.002 & 0.002 & 0.002 & 0.002 & 0.002 & 0.004 & 0.004 & 0.005 \\ 
 Online Fixed ModelSel   & 0.004 & 0.004 & 0.002 & 0.001 & 0.001 & 0.001 & 0.001 & 0.002 & 0.003 & 0.004 \\ 
   \hline
\end{tabular}
}
\caption{False Positive Rates for Offline, Online Changing ModelSel, and Online Fixed ModelSel for Simulation Scenario 2 for batches $11,15, \dots, 47$. The values are averaged over 25 replications.}
 \label{tab:sim2:fpr}
\end{table}

\subsubsection{Simulation Scenario 3}
\begin{figure}[ht]
    \centering
    \includegraphics[height=4.4in,width=4.8in]{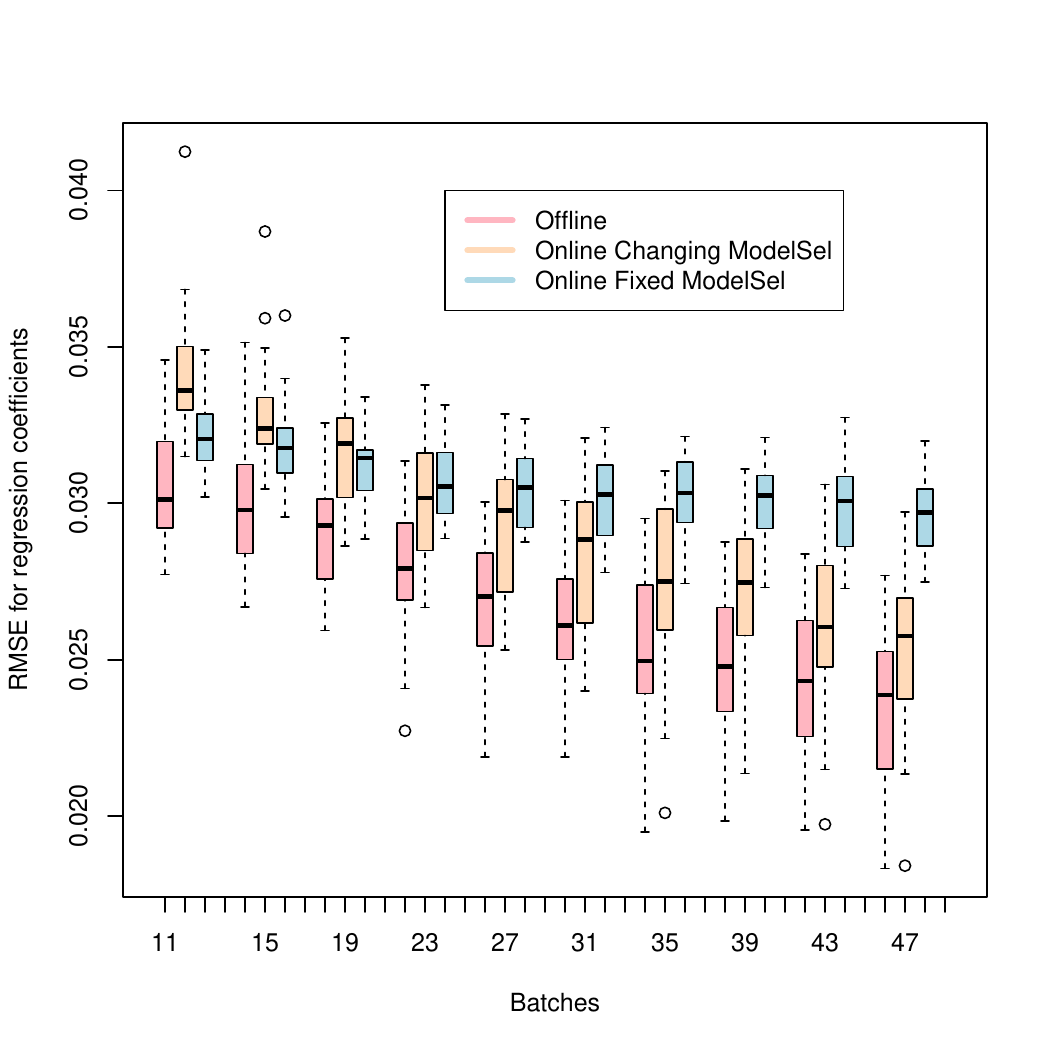} 
    \includegraphics[height=4.4in,width=4.8in]{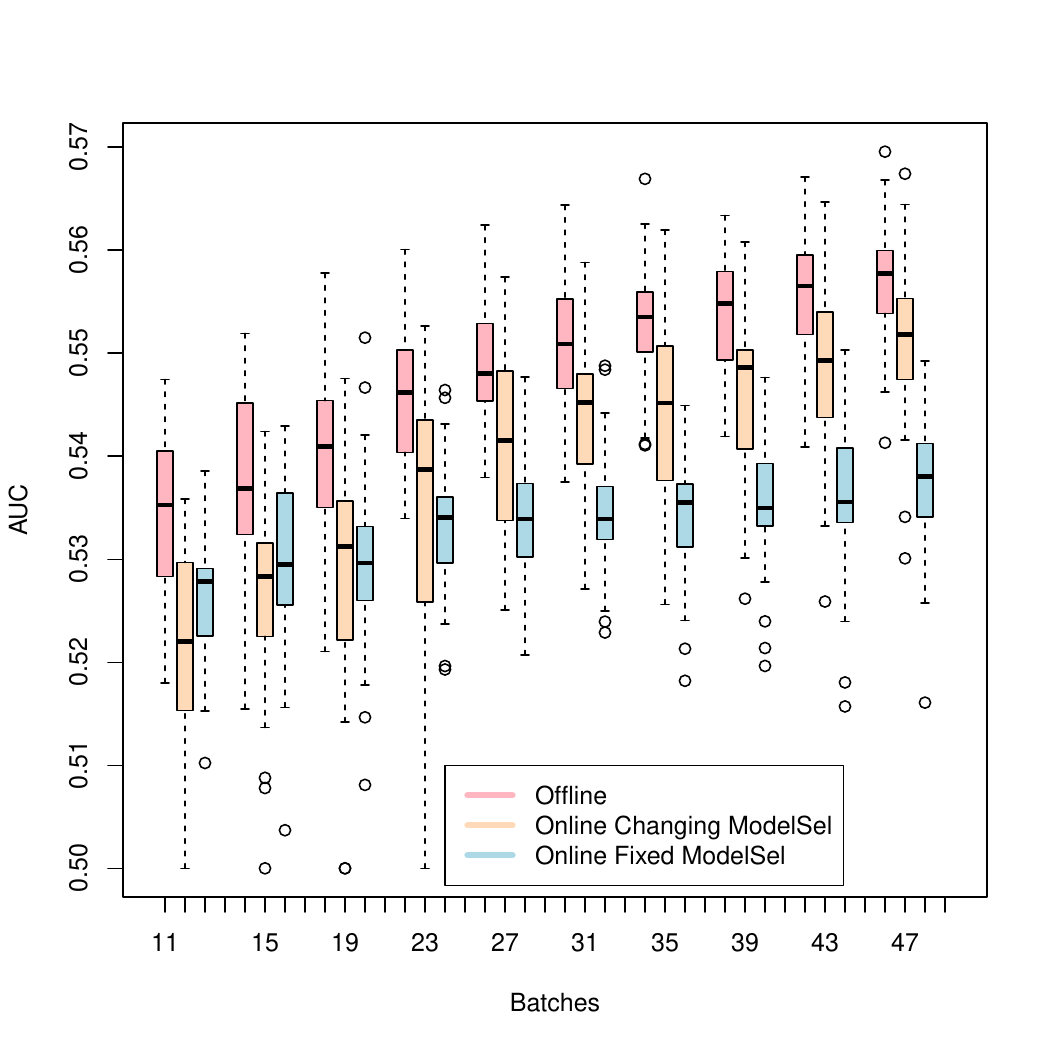}
    \caption{RMSE of regression coefficients and AUC for test data
    for Offline, Online Changing ModelSel, and Online Fixed ModelSel for Scenario 1. The boxplots are based on 25 replicates. \textbf{The running times of Offline and Online Fixed ModelSel are approximately 344 and 7 times that of Online Changing ModelSel for a single replication.}}
    \label{fig:3}
\end{figure}

Finally for the third choice of $\beta=(1/1.5)\times(0.1,\dots,0.1,0,\dots,0)^T$,
where the signals are weaker, the differences between the methods become more pronounced. Figure \ref{fig:3} shows that Offline has the strongest performance, followed by Online Changing ModelSel, which in turn is followed by Online Fixed ModelSel. The weaker signals make it challenging for all methods to consistently detect true variables, as indicated by
the lower TPR in Table \ref{tab:sim3:tpr}. Online Fixed ModelSel
struggles the most in this scenario, displaying higher RMSE in Figure \ref{fig:3} and a greater tendency to miss relevant variables in Table \ref{tab:sim3:tpr}. The missed variables also lead to an overall lower AUC
for this method, as evident from Figure \ref{fig:3}. While Offline and Online Changing ModelSel outperform in this scenario by accurately identifying the true variables, even these methods face considerable difficulty due to the weak signals resulting in a lower AUC for all methods compared to Scenarios 1 and 2.

\begin{table}[ht]
\centering
\footnotesize{
\begin{tabular}{|c|cccccccccc|}
  \hline
   & 11 & 15 & 19 & 23 & 27 & 31 & 35 & 39 & 43 & 47 \\
  \hline
Offline & 0.084 & 0.124 & 0.158 & 0.222 & 0.264 & 0.304 & 0.338 & 0.362 & 0.398 & 0.444 \\ 
  Online Changing ModelSel & 0.084 & 0.128 & 0.156 & 0.218 & 0.262 & 0.304 & 0.340 & 0.354 & 0.398 & 0.432 \\ 
  Online Fixed ModelSel  & 0.058 & 0.074 & 0.076 & 0.096 & 0.110 & 0.148 & 0.140 & 0.168 & 0.162 & 0.168 \\ 
   \hline
\end{tabular}}
\caption{True Positive Rates for Offline, Online Changing ModelSel, and Online Fixed ModelSel for Simulation Scenario 2 for batches $11,15, \dots, 47$. The values are averaged over 25 replications.}
 \label{tab:sim3:tpr}
\end{table}  

\begin{table}[ht!]
\centering
\footnotesize{
\begin{tabular}{|c|cccccccccc|}
  \hline
    & 11 & 15 & 19 & 23 & 27 & 31 & 35 & 39 & 43 & 47 \\
  \hline
Offline & 0.007 & 0.008 & 0.006 & 0.004 & 0.003 & 0.003 & 0.003 & 0.002 & 0.003 & 0.002 \\ 
  Online Changing ModelSel & 0.009 & 0.008 & 0.006 & 0.005 & 0.002 & 0.003 & 0.003 & 0.001 & 0.003 & 0.002 \\ 
   Online Fixed ModelSel & 0.004 & 0.005 & 0.003 & 0.003 & 0.003 & 0.003 & 0.002 & 0.002 & 0.002 & 0.002 \\ 
   \hline
\end{tabular}}
\caption{False Positive Rates for Offline, Online Changing ModelSel, and Online Fixed  ModelSel for Simulation Scenario 2 for batches $11,15, \dots, 47$. The values are averaged over 25 replications.}
 \label{tab:sim3:fpr}
\end{table}

\subsection{Comparison of Running Time and Summary of Results}

In real-world applications, computational time impacts the feasibility and scalability of methods, especially with large datasets or complex models. An efficient method that achieves accurate results in less time can significantly reduce computational costs and enhance productivity. Therefore we consider the execution time comparison below, which is crucial for streaming data. The time taken for 1 replication of 50 batches with 10,000 MCMC iterations after a 
2,000 burn-in for each method is recorded in the following table:

\begin{table}[h!]
\centering
    \begin{tabular}{|c|c|c|c|c|}
    \hline
        & Scenario 1 & Scenario 2 & Scenario 3 \\
    \hline    
        Offline &  4.04 hours & 4.35 hours & 3.06 hours\\
        & (441) & (489) & (344)\\
        \hline
        Online Fixed ModelSel & 8.77 minutes & 6.25 minutes &  3.49 minutes\\
        & (16) & (12) & (7)\\
        \hline
        Online Changing ModelSel &  33 seconds &  32 seconds &  32 seconds\\
    \hline    
    \end{tabular}
    \caption{Time needed for a single replication for the three compared methods in the three simulation scenarios. The numbers in parentheses denote
    the running time of the method relative to that of Online Changing ModelSel; for example, for Scenario 1, Offline and Online Fixed ModelSel take 441 and 16 times that of Online Changing ModelSel, for one replicate.}
    \label{tab:sims:time}
\end{table}

In this simulation study the Online Changing ModelSel method consistently outperforms Online Fixed ModelSel and more closely matches the ideal results from Offline, for larger sample sizes. Furthermore, Online Changing ModelSel offers significant time savings, depicted in Table \ref{tab:sims:time}, without compromising accuracy that much, making it an efficient method for real-time Bayesian model selection. Offline provides reliable results with minimal variability, while Online Fixed ModelSel shows increased variability and less stability, particularly with weaker signals, when important variables that are dropped in the screening stage cannot be recovered easily. Overall, when Offline is prohibitively slow,
Online Changing ModelSel provides a useful alternative to
Online Fixed ModelSel, with faster results and greater accuracy at large sample sizes.

\section{Application of the Methods to Traffic Crash Data}
\label{sec:real}
We have applied the established methodologies to the streaming data from the National Automotive Sampling System (NASS) Crashworthiness Data System (CDS), focusing on occupant-level information from 2009 to 2015 used in \cite{ghos:tan:luo:2020}. Our analysis centers on whether occupants sustained injuries during traffic crashes, using the MAIS (maximum abbreviated injury scale) as a key indicator (Luo and Song, 2020). 
Predictors include occupant characteristics such as AGE, HEIGHT, OCCRACE (race), OCETHNIC (ethnicity), ROLE (driver/passenger), SEX, WEIGHT, alongside vehicle and accident-related factors: BAGFAIL (air bag system failure), BELTANCH (shoulder belt anchorage adjustment), EJCTAREA (ejection area), and PARUSE (restraint use, as reported by police).
The dataset has 33 predictors, with 300 noise variables added, as in \cite{ghos:tan:luo:2020}. 

For Offline and Online Fixed ModelSel (referred to as Online 2 in \cite{ghos:tan:luo:2020}), we reproduced the analysis in \cite{ghos:tan:luo:2020}, with 20 batches, each of size 800, and kept the last 438 observations as test data as in \cite{ghos:tan:luo:2020}. The MPM from the Offline method includes AGE, a SEX category, WEIGHT, all the PARUSE categories, and one EJCTAREA category. Online Fixed ModelSel uses the first 5 batches for screening, and switches to online estimation from the 6th batch. It retains mostly the same variables as Offline in its MPM; however it drops one PARUSE category related to shoulder belt use, which is included by Offline. 

For the Online Changing ModelSel method, we use the same model space priors  as \cite{ghos:tan:luo:2020}, that is we use a Beta(1,1) prior on the inclusion probability, instead of fixing it at 0.5, because the model
space is relatively large with 333 predictors, and sparsity can be better promoted through a Beta-Binomial prior on the model space, than a discrete uniform prior, used in the simulation studies.
Moreover, unlike the simulation studies, this dataset exhibits separation, so that the MLE does not exist for the first batch. A proper prior is required for the regression coefficients in order for the posterior mode to exist. \cite{ghos:tan:luo:2020} used the 
log $F(m,m)$ prior \citep{Gree:Mans:2015}, only
on the variables that are related to separation.
For the Online Changing ModelSel method, we decide to use 
independent normal priors on all the regression coefficients instead, which can offer adequate
shrinkage under separation. The choice of independent normal priors has been shown to 
perform well by \cite{Ghos:Li:Mitr:2018} under separation. Since we approximate the estimate under the MPM, by thresholding coefficients under the full model, using a prior that leads to adequate shrinkage under separation is particularly important. We use the approximate EM algorithm by
\cite{Gelm:Jaku:Pitt:Su:2008} based on their 
R package {\tt arm}, to calculate the approximate posterior mode under independent normal priors, 
during the first 5 batches, which amounts to the screening stage of Online Fixed ModelSel, when all individual records are assumed to be available for all methods. From the sixth batch, we update the modal estimates using the method of \cite{luo:song:2020} using their R package {\tt RenewGLM}, for online updating without access to individual records of past batches.

\begin{figure}[ht]
    \centering
    \includegraphics[height=5in,width=5.5in]{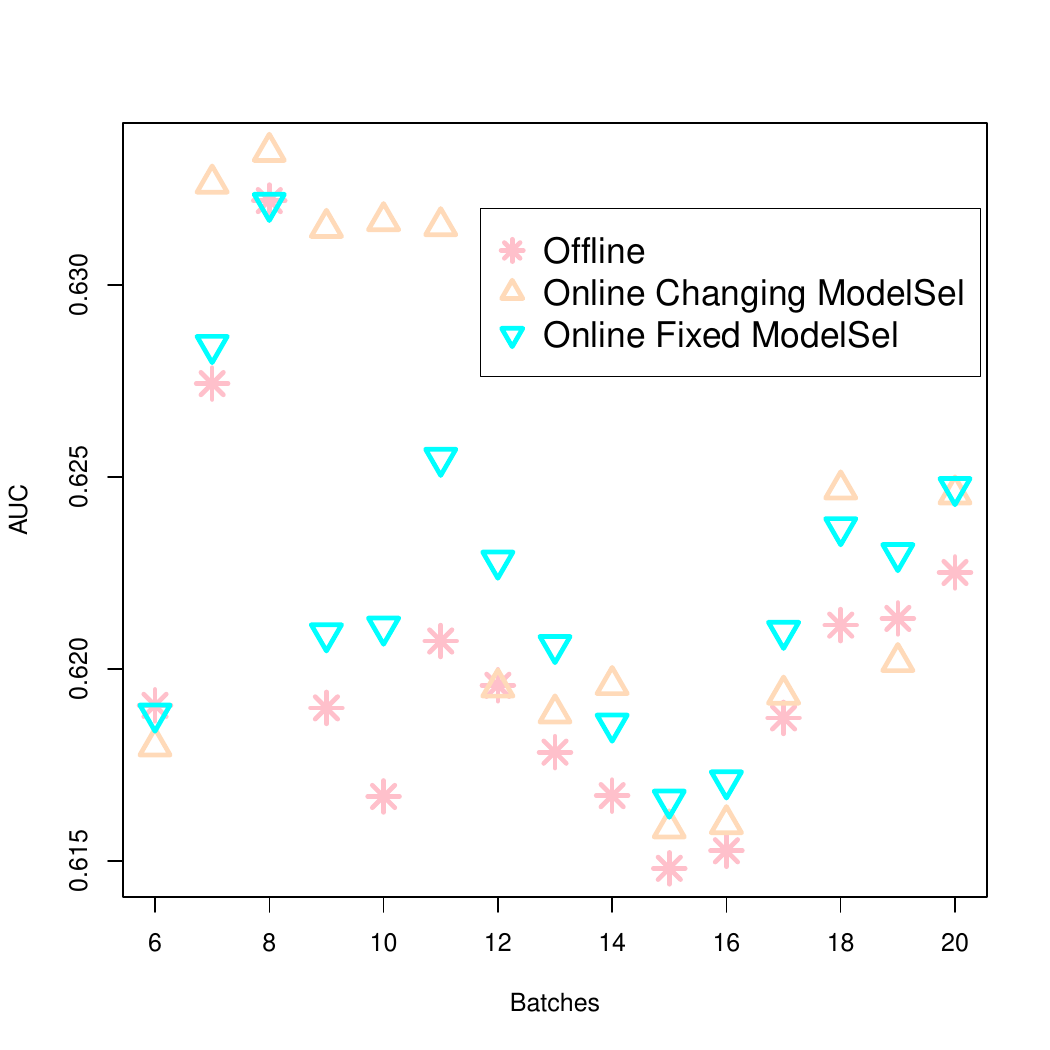}
    \caption{AUC for Offline, Online Changing ModelSel, and Online Fixed ModelSel for the traffic crash test data.
     \textbf{The running times of Offline and Online Fixed ModelSel are approximately 225 and 12 times that of Online Changing ModelSel for this data.}}
    \label{fig:4}
\end{figure}

Interestingly our Online Changing ModelSel includes identical variables as Offline, and is able to identify the variable that was missed by Online Fixed ModelSel. The running times for Offline, Online Fixed ModelSel, and Online Changing ModelSel are 15 hours, 49 minutes, and 4 minutes approximately. Therefore our new Online Changing ModelSel method outperforms Online Fixed ModelSel in variable selection while also being around 11 times faster in execution. The values of AUC in a test dataset for all methods are given in Figure \ref{fig:4}.
No method has uniformly higher AUC values across all batches, with Online Changing ModelSel being competitive with the existing Online Fixed ModelSel method.

\section{Discussion}
\label{sec:summary}
The Online Changing ModelSel method proposed in this paper is a promising method for online variable selection in logistic regression, particularly suited for high-dimensional streaming data. It is a challenging task to perform Bayesian model selection in logistic regression for streaming data, in an online setting without access to individual subject level information on historical data. 
The proposed method is based on a linear model approximation
to the logistic regression model, which makes it computationally feasible to implement online MCMC algorithms
that allow selected models to be updated or changed as new batches of data are collected. Furthermore, the online MCMC
gives mathematically identical results to offline MCMC under a linear regression set up. Empirical results suggest that the proposed method can be quite effective in variable
selection, with results that are very close to offline Bayesian model selection for both simulated and real data. 

The proposed method estimates the regression coefficients in the MPM, by thresholding the Bayesian full model estimates, according to the selected variables in the MPM.
Since the offline and Online Fixed ModelSel method of 
\citet{ghos:tan:luo:2020} use BMA estimates instead, in some situations the proposed method could have higher RMSE
in estimating regression coefficients for initial batches.
However, as the sample size gets larger in later batches,
its performance is similar or better than the method of \citet{ghos:tan:luo:2020}, across all scenarios. 

In the simulation studies, the Online Changing ModelSel method approximated the posterior mode of the regression coefficients by the MLE, based on a large sample
assumption. For the real data, a normal prior was used to deal with separation, as the MLE does not exist for the first batch. It is possible that using the normal prior in the simulation studies would enhance the performance of the 
Online Changing ModelSel method further, by shrinking the
online Bayesian full model estimates further to zero, compared to the MLE. In this article, our focus has been on Bayesian model selection for logistic regression, but the method can be extended to other GLMs such as Poisson regression.

\bibliography{online.bib, jcgs-online.bib}







\end{document}